\documentclass[usegraphicx,usenatbib]{mn2e}
\usepackage{epsf}

\def\omh{\Omega_{\rm m} h}
\def\omb{\Omega_{\rm b}}
\def\om{\Omega_{\rm m}}

\def \bj {b_{\rm J} }

\begin{document}

\title[Precision Cosmology from Large Scale Structure?]
{
The Galaxy Power Spectrum: Precision Cosmology from Large Scale Structure?
}
\author[Ariel S\'anchez \& Shaun Cole]
{\parbox[t]{\textwidth}{
Ariel G. S\'anchez$^{1,2}$\thanks{E-mail: arielsan@oac.uncor.edu},
Shaun Cole$^{2}$.
}
\vspace*{6pt} \\ 
$^{1}$ Instituto de Astronom\'{\i}a Te\'orica y Experimental (IATE), OAC, UNC, 
Laprida 854, X5000BGR, C\'ordoba, Argentina.\\
$^{2}$ The Institute for Computational Cosmology, 
Department of Physics, University of Durham, South Road, Durham DH1
3LE, UK.\\
\\
}
\date{Submitted to MNRAS}
\maketitle

\begin{abstract}
Published galaxy power spectra from the 2dFGRS and SDSS
are not in good agreement. We revisit this issue by
analyzing both the 2dFGRS and SDSS DR5 catalogues
using essentially identical techniques. We confirm
that the 2dFGRS exhibits relatively more large scale power
than the SDSS, or, equivalently, SDSS has more small scale 
power. We demonstrate that this difference is due to
the $r$-band selected SDSS catalogue being dominated by
more strongly clustered red galaxies, which have a stronger
scale dependent bias. The power spectra of galaxies
of the same rest frame colours from the two surveys
match well. If not accounted for, the difference between the SDSS and
2dFGRS power spectra causes a bias in the obtained constraints on cosmological
parameters which is larger than the uncertainty with which they are 
determined. We also found that the correction developed by \citet{cole05} to model
the distortion in the shape of the power spectrum due to non-linear evolution and 
scale dependent bias is not able to reconcile the constraints obtained from the 2dFGRS
and SDSS power spectra. 
Intriguingly, the model is able to describe the differences between the 2dFGRS and the much 
more strongly clustered LRG sample, which exhibits greater nonlinearities.
This shows that more work is needed to understand the relation between the galaxy power spectrum and the
linear perturbation theory prediction for the power spectrum of matter fluctuations.
It is therefore important to accurately model these effects to get precise 
estimates of cosmological parameters from these power spectra and from future 
galaxy surveys like Pan-STARRS, or the Dark Energy Survey, which will use selection 
criteria similar to the one of SDSS.

\end{abstract}


\section{Introduction}

In the last decade, the advent of new precise cosmological observations coming mainly from 
measurements of fluctuations in the temperature of the cosmic microwave background radiation (CMB) and 
the large scale structure of the Universe (LSS), have shown a dramatic improvement. This has revolutionized
our knowledge of the values of the basic cosmological parameters which are constrained to an accuracy of around 10\% \citep{sanchez06,spergel07}.

In particular, the measurements of large scale galaxy clustering place important constraints on cosmological
parameters that complement those from the analysis of fluctuations in the CMB. Measurements of the galaxy
power spectrum constrain the parameter combinations $\omh$ and $\omb/\om$. The constraint on $\omh$ is
particularly important as, for instance, it breaks a degeneracy in the CMB data and allows accurate determination of $\om$ \citep[e.g.][]{efstathiou02}.

In view of these rapid improvements in the amount and quality of the observations
the control of the systematic effects introduced in the analysis pipeline becomes increasingly important.
New experiments and surveys are being planned which will push the level of precision of the constraints on cosmological parameters even further. Thus it is necessary to establish how robust the constraints really
are with respect to the details of the hypothesis implemented to establish the link between theory and observations.

\begin{figure*}
\centering
\centerline{\includegraphics[width=0.85\textwidth]{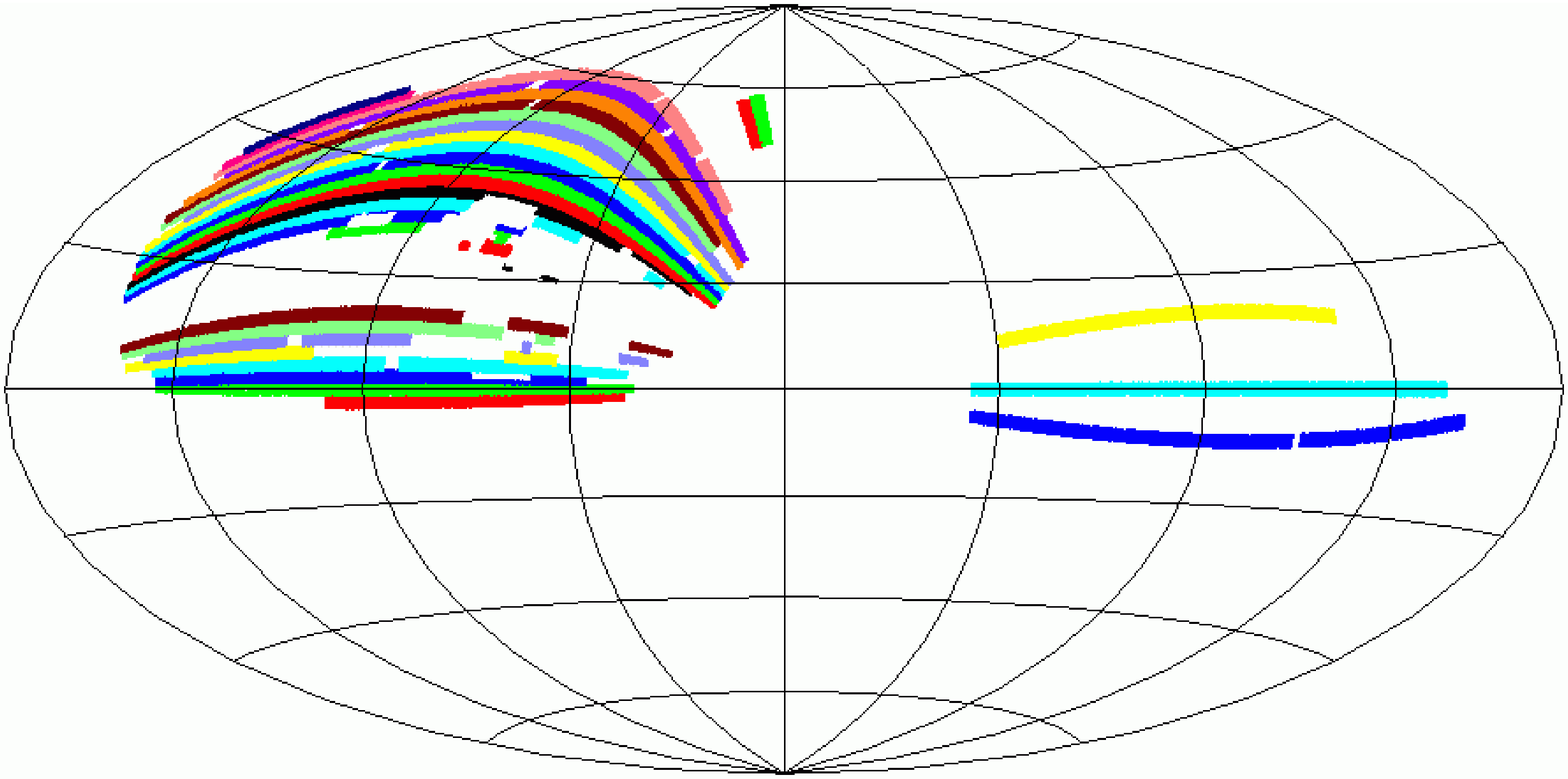}}
\centerline{\includegraphics[width=0.85\textwidth]{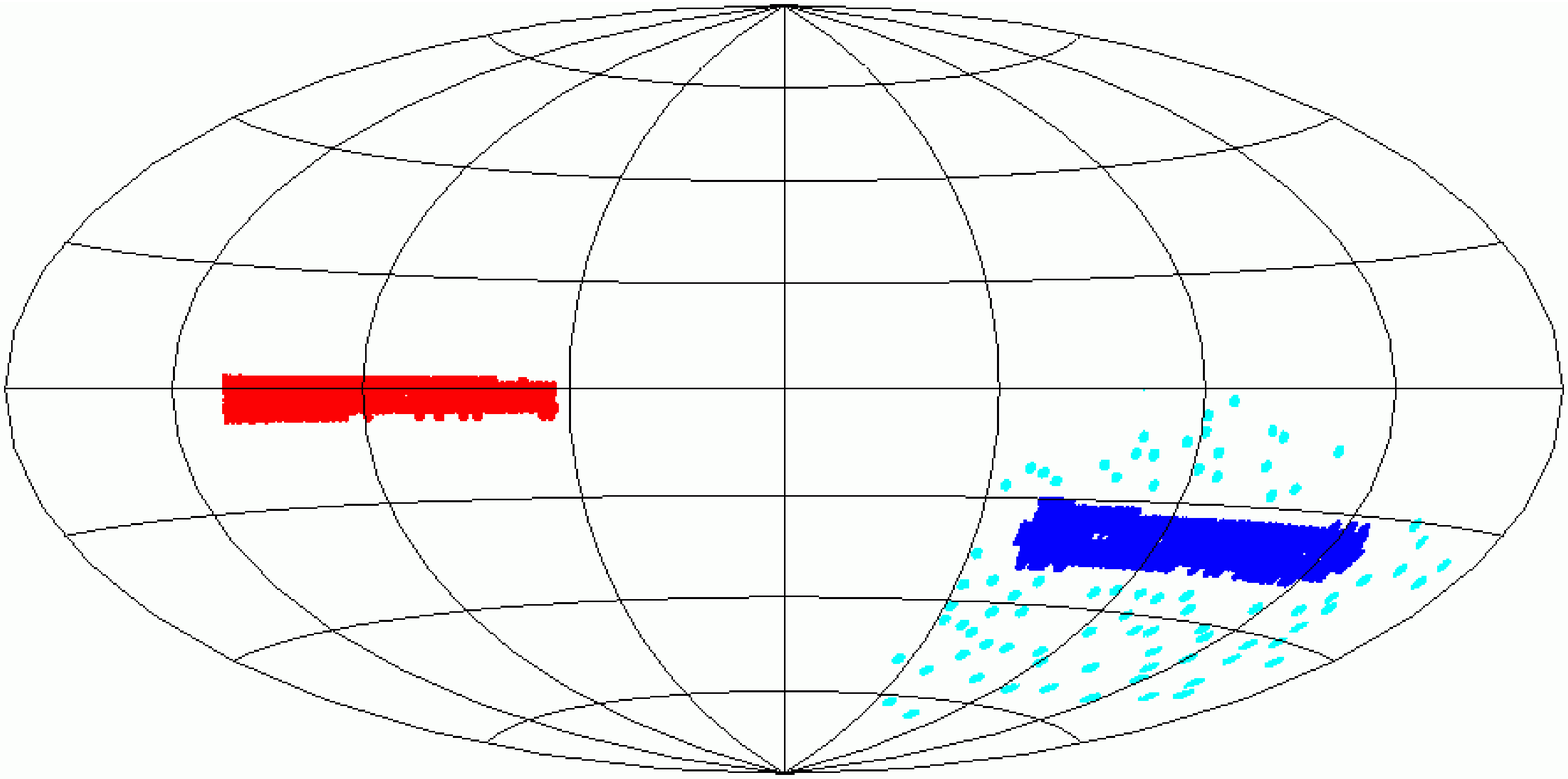}}
\caption{
Equal area, all sky, Aitoff projections of all the
galaxies accepted by our SDSS (upper) and
2dFGRS (lower) masks used in our analysis.
}
\label{fig:mask}
\end{figure*}

The tension found between the constraints coming from the published analysis of the two-degree field galaxy 
redshift survey (2dFGRS) of \citet{cole05} and the Sloan Digital Sky Survey (SDSS) of \citet{tegmark04} might be an indication of possible systematic effects. In figure~16a of
\citet{cole05}, which compares the published estimates of the galaxy
power spectra, there is evidence of more large scale power in
the 2dFGRS than in SDSS. This folds through and results in the
headline value of $\omh =0.168\pm 0.016$ from 2dFGRS \citep{cole05}
being lower than that of \citet{tegmark04} SDSS, $\omh = 0.213\pm 0.023$.
Here, one should be cautious as different priors have been assumed,
but in the analysis of \citet{sanchez06}, which treats each dataset on
an equal footing and separately combines each with CMB data, 
one sees in their figure~18 that the SDSS prefers a substantially
higher value of $\om$ than does the 2dFGRS. In fact, while the
2dFGRS estimate is in good agreement with that from the CMB data alone
(it significantly tightens the constraint without shifting the
best fitting value) the SDSS data pull $\om$  to higher
values than preferred by the CMB.

The 2dFGRS and SDSS are the largest redshift surveys available and give
the most detailed description of the large scale structure of the Universe 
as traced by galaxies. Hence to be sure
that systematic errors are not biasing the parameters it is very
useful to see each data set subjected
to analysis by a variety of algorithms and codes, as happened
following the \citet{percival01} 2dFGRS power-spectrum analysis
\citep{tegmark01}.

\citet{percival07a} computed the power spectrum of a combined Main galaxy
and Luminous Red Galaxy (LRG) sample drawn from SDSS Data Release 5 using a
similar technique to that of \citet{cole05} and found an even higher value of
the matter density $\Omega_mh=0.23 \pm 0.01 $, suggesting that the inclusion
of the LRGs in the analysis increase the discrepancy between 2dFGRS and SDSS.
Their results also show that if the analysis is restricted to large scales 
($0.01<k<0.06\,h~\mathrm{Mpc}^{-1}$) the data favours a lower matter density
$\Omega_mh=0.16 \pm 0.03 $. \citet{percival07a} suggested that these differences
could be explained by
scale-dependent galaxy bias on large scales, but found no
significant evidence of it.

Here we seek to investigate whether these differences
are as a result of: a)~larger than expected cosmic variance, 
b)~systematics due to differences in the analysis technique 
c)~systematics due to problems with galaxy catalogues or d) intrinsic
differences in the underlying galaxy clustering.  In order to directly
compare the 2dFGRS and SDSS, we analyse each dataset using essentially
identical methods which we outline in Section~\ref{sec:method}.  In
Section~\ref{sec:overlap}, we briefly look at the region of overlap
between the surveys to study the different selections used
and the level of incompleteness. In Section~\ref{sec:pkcomparison}, we
compare the resulting power spectra from the full catalogues and interpret the differences.
Section~\ref{sec:pkconstraints} presents a detailed analysis of the implications of
the differences between these datasets in the obtained constraints on cosmological parameters, and the systematic effects that might be introduced by the analysis.
Finally, we conclude in Section~\ref{sec:conc}

\section{Methods}
\label{sec:method}

 \subsection{2dFGRS analysis}   

The 2dFGRS covers approximately 1800 square degrees distributed
between two broad strips, one across the South Galactic Pole (SGP)
and the other close to the North Galactic Pole (NGP), plus a set
of 99 random 2 degree fields  spread over the full southern galactic cap. 
The final catalogue contains reliable redshifts for 221\,414 galaxies selected
to an extinction-corrected magnitude limit of approximately $b_{\mathrm{J}} = 19.45$ 
\citep{colless01,colless03}. The 2dFGRS samples analysed here are the same as described in detail in \citet{cole05}.

  Our method of estimating the galaxy power spectrum and determining
statistical errors is essentially identical to that set out
in \citet{cole05}, but with two minor changes. In brief:

We use masks, whose construction is described in 
\citet{norberg02}, to describe the angular variation of the survey magnitude
limit, redshift completeness and magnitude dependence of the redshift
completeness.

Random catalogues are generated by sampling from the
luminosity function and viewing through the masks. To generate
random catalogues corresponding to red/blue subsets, the luminosity
function of only the red/blue galaxies is used.

Close pair incompleteness due to ``fibre collisions''
is dealt with by redistributing the weights of missed galaxies to their
10 nearest neighbours on the sky.

The power spectrum is estimated using a simple cubic 
FFT method with the optimal weighting scheme of \citet[hereafter PVP]{PVP} and then
spherically averaged in redshift space. As in \citet{cole05}, the assumed linear 
empirical bias factors that are used in this weighting scheme are
\begin{equation}
b_{\rm blue} =0.9\, (0.85+0.15\, L/L_*),
\label{eqn:bias}
\end{equation}
for rest frame $b_{\rm  J} -r_{\rm F} <1.07$, and
\begin{equation}
b_{\rm red} = 1.3\, (0.85+0.15\, L/L_*),
\label{eqn:bias_red}
\end{equation}
for rest frame $b_{\rm  J} -r_{\rm F} >1.07$.
These are based on the scale-independent bias parameter found by Norberg et al. (2001), modified
to describe the difference in amplitudes of the power spectra of red and blue galaxies around
$k=0.1\,h\,{\mathrm{Mpc}}^{-1}$. 
This simple description of the luminosity and colour dependence of galaxy bias is in agreement with the results of \citet{swanson07}.

\citet{cole05} showed that the effect on the recovered power spectrum
of using \citet[hereafter FKP]{FKP} rather than the PVP estimator is small
(see their figures~17(s) and (t)).
The FKP estimator is biased, as it ignores the luminosity dependence of galaxy clustering.
Provided the model of luminosity-dependent bias is correct, then the PVP estimator removes this bias.
A more complicated recipe for the luminosity and colour dependence of galaxy bias can be
applied but, as long as the assumed bias is scale independent (a necessary assumption in our method of
analysis) it will hardly change the shape of the estimated power spectrum, as the effect will be significantly smaller than the difference between the FKP and PVP estimates and so entirely negligible.

The covariance matrix describing the errors
on the power spectrum measurements and their correlations is 
estimated using mock catalogues which are constructed from the 
random catalogues by generating a log-normal density field with
a specified power spectrum
and using it to modulate the selection of galaxies from the
random catalogue. 
Thus, by construction, these catalogues have a power spectrum very
close to the best fitting model and have luminosity and colour
dependent clustering consistent with the bias factors of
equations~(\ref{eqn:bias}) and~(\ref{eqn:bias_red}).

The survey window function is determined directly from
the random catalogue. When fitting models the theoretical model power
spectra are convolved with the survey window function.

The two minor changes we have made are:
\begin{itemize} 
    \item{} We have adopted a simpler binning scheme so that now $P(k)$
is estimated in bins uniformly spaced in $\log_{10} k$, rather
than the linearly space bins with different bin widths in different
ranges of $k$ that were used in \citet{cole05}.
    \item{} We used new sets of log-normal catalogues in which 
the modulation of the density field used for galaxies with different 
bias factors is linear rather than the slightly more complicated 
scheme that was employed by \citet{cole05}.
\end{itemize} 

\begin{figure}
\includegraphics[width=0.49\textwidth]{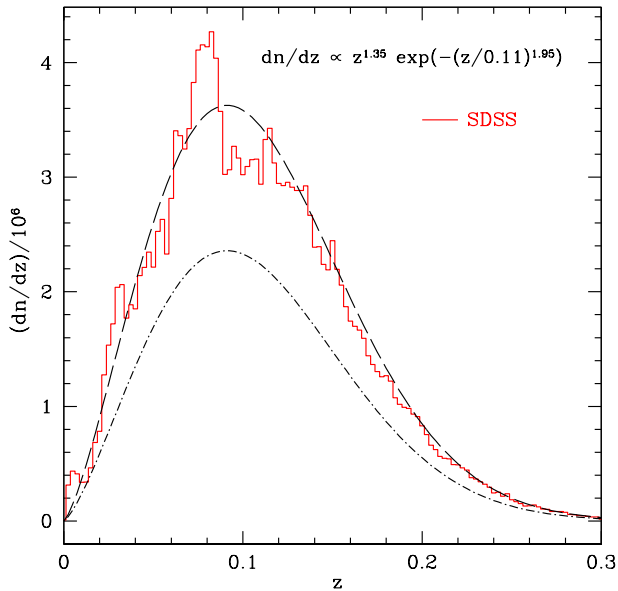}
\caption{
The redshift distribution of the SDSS $r<17.77$ galaxy sample (solid line).
The dashed line shows an analytic fit to the data. The dot-dashed line is this same 
fit scaled down in amplitude to be always below the redshift histogram. This fit is
used to generate the corresponding random catalogue by sampling from the original survey
as described in the text.
}
\label{fig:dndz}
\end{figure}

 \subsection{SDSS analysis}   

We have analysed the SDSS Data Release 5 (DR5) sample \citep{dr5}, which is considerably larger than 
the DR3 sample analysed by \citet{tegmark04}.
In most respects our analysis of the SDSS is identical to that of
the 2dFGRS. The only differences are a simpler way of generating 
the survey masks and of populating the corresponding random catalogue.

The sky coverage mask we have 
adopted for the SDSS-DR5 data
is shown in Fig.~\ref{fig:mask} and compared with corresponding
2dFGRS mask. We constructed this mask by simply noting the 
angular coverage of each of the stripes from which the SDSS
survey is built and by removing a few small regions with poor
coverage. Most of the SDSS survey goes to a uniform magnitude limit of
$r=17.77$, but a sub-area, which is easily identified using the
target selection date, has a variety of different magnitude limits.
In this sub-area we simply imposed a fixed magnitude limit of
$r=17.5$ and discarded all galaxies fainter than this limit.
The number of galaxies with redshifts that are retained
by the mask and magnitude limits and used in our analysis is $443\, 424$.
The mask is cruder than
the more sophisticated ones employed by \citet{tegmark04} and for
the 2dFGRS as it ignores the smaller scale variation in the redshift
completeness. However in the case of the 2dFGRS, where the
incompleteness variation is much larger, \citet{cole05} showed that
provided this incompleteness is accounted for using our method of
redistributing the weights of galaxies without redshifts to
neighbours with redshifts, the resulting power spectrum estimates
are very accurate (see their figure~17g). 

\begin{figure*}
\centering
\centerline{\includegraphics[width=0.95\textwidth]{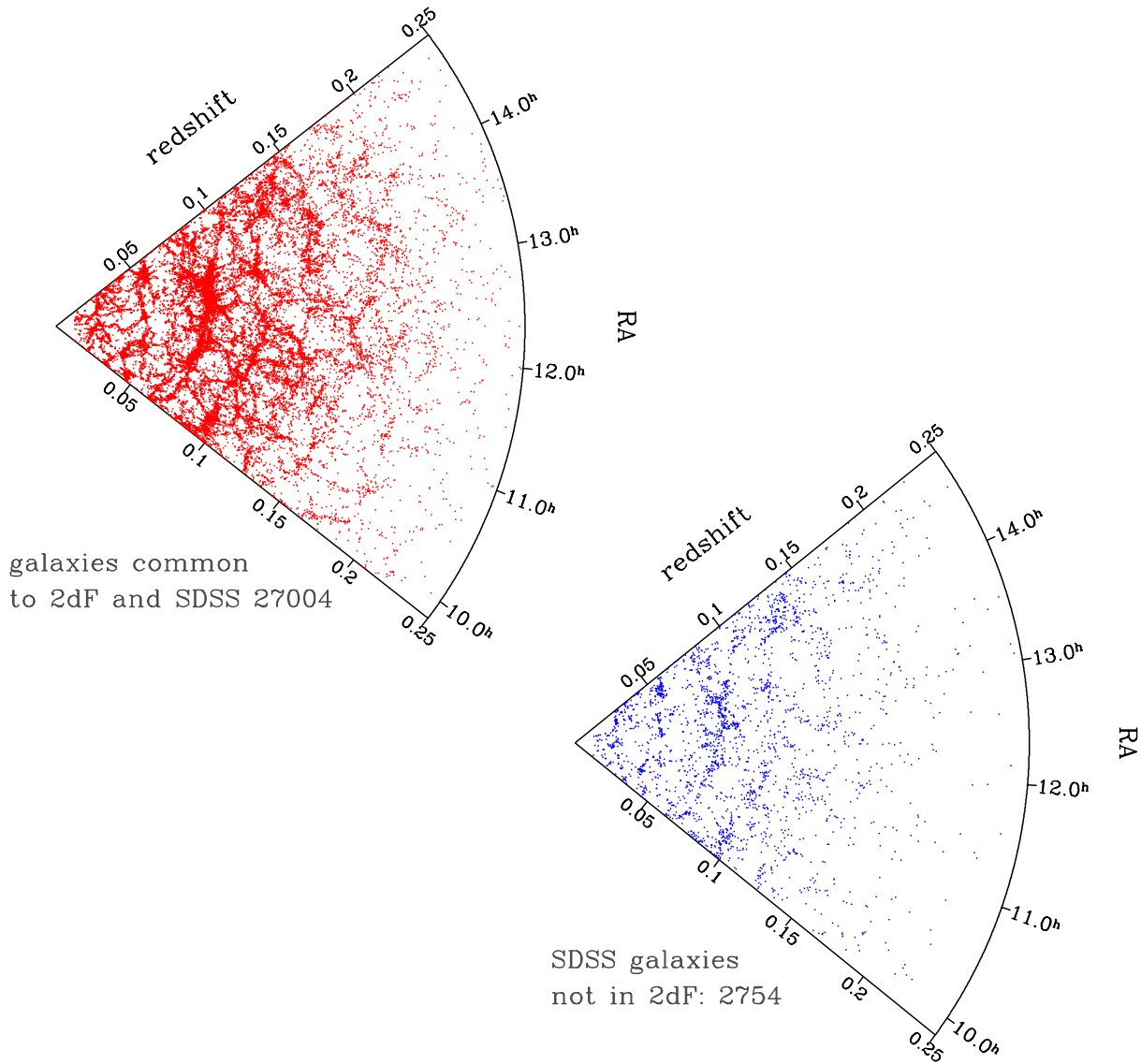}}
\caption{
Cone plots showing RA and redshift for galaxies 
in the region of sky where the 2dFGRS and SDSS surveys overlap.
There are $53\, 382$  within this area that are in both
surveys and in the upper panel we plot the sub-sample of  $27\, 004$
that (estimated from SDSS photometry) have $b_{\rm J}<19$
and redshift $z>0.01$. 
The lower panel shows the $2754$ galaxies that are in SDSS and pass
the same magnitude and redshift cuts, 
but are missing from the 2dFGRS catalogue.
}
\label{fig:cones}
\end{figure*}

As most of the catalogue goes uniformly to the deeper
magnitude limit a simple method can be used to  construct the
random catalogue. Fig.~\ref{fig:dndz} shows the redshift distribution
of this sample (solid line) together with an analytic fit (dashed line) 
that smoothes away the effect of large scale clustering. Our procedure is:
\begin{enumerate}
\item{}Select a random direction on the sky.
\item{}Choose at random a genuine galaxy from the region of
  the survey that goes to $r=17.77$ .
\item{}Keep the galaxy with a probability proportional to the ratio
of the fit scaled down in amplitude
to be always below the redshift histogram (shown as the dot-dashed line in
Fig.~\ref{fig:dndz}) to that of the actual height of the redshift histogram
at the redshift of the selected galaxy. This effectively adjusts the sampling of
the different redshifts to erase the effect of large scale clustering.
\item{}Keep or discard the galaxy based on the sky coverage and
magnitude limit of the mask. (Note, for galaxies that fall where
the  magnitude limit is only $17.5$, the fainter
galaxies will be discarded and the redshift distribution of the
retained galaxies will be appropriately shallower than that of
Fig.~\ref{fig:dndz}.)

\end{enumerate}

These steps are done repeatedly until a random catalogue
containing $100$ times more galaxies than the genuine catalogue
is built up. Selecting from the genuine catalogue in this way
means we automatically have apparent magnitudes and colours for
all the galaxies in the random catalogue and so can select sub-samples
from it and weight its galaxies in just the same way as the genuine
catalogue.

To utilize the \citet{PVP} optimum weighting we need to 
determine bias factors for the galaxies in 
the genuine, random and mock catalogues.
We do this  by first converting the SDSS
magnitudes to the 2dFGRS $b_J$ and $r_F$ bands using 
\begin{eqnarray}
b_{\rm J} &=& g + 0.15 +0.13\, (g-r) \cr
r_{\rm F} &=& r -0.13
\label{eqn:col}
\end{eqnarray}
and the simple colour dependent $k$-corrections that were
used for the 2dFGRS data \citep[see section~3 of][]{cole05}.
Then we are able to define the bias factors using
equations~(\ref{eqn:bias}) and~(\ref{eqn:bias_red}) just as for the 2dFGRS data.


\section{Survey overlap}
\label{sec:overlap}
Our main focus is the comparison of the power spectra of the
two surveys, but we first directly compare the two surveys
in the region of their overlap to get a feel for the different
selections used and the level of incompleteness.

In the northern galactic hemisphere there is a contiguous area of
overlap between the two surveys, which runs for $74$~degrees of RA
and is for the most part $5.2$~degrees wide in declination. 
If we select from the SDSS photometric catalogue all galaxies
brighter than $\bj=20$ (we do not apply the $r \approx 17.77$
magnitude limit of the SDSS main galaxy survey, but we do apply
all the other star-galaxy classification criteria used in that sample
\citep[see][]{strauss02}), then in this area there are $53\, 382$ galaxies that are in 
both catalogues. We find the fraction of SDSS galaxies which are
also in the 2dFGRS to be constant at $89$\% as faint as $\bj \approx
18.9$. Fainter than this, SDSS galaxies are missing from the 2dFGRS
sample simply due to the (variable) magnitude limit of the 2dFGRS
survey and its $0.15$~magnitude random photometric errors.
This finding is in perfect accord with the estimates 
made with the SDSS EDR \citep{stoughton} in \citet{norberg02}.
This $11$\% incompleteness has been investigated by
\citet{cross} as well as \citet{norberg02} and has been shown to be
predominately due to incorrect star-galaxy classification.
The star-galaxy classification parameters based on the
APM photometry are noisy and this level of 
incompleteness  is in line with what was expected \citep{maddox90}.

The issue here is whether this incompleteness has any influence
on estimates of galaxy clustering. We can look at this
directly by plotting cone plots (Fig.~\ref{fig:cones}) of the
galaxies the two catalogues have in common and those missed
by the 2dFGRS. We plot only galaxies with $\bj<19$ to avoid
issues with the 2dFGRS magnitude limit.
We see that $91$\% of the SDSS galaxies are in the 2dFGRS. 
Comparing the two cone plots in Fig.~\ref{fig:cones}, it
appears that the galaxies missed by 2dFGRS are just a random
sparse sampling of the structure seen in the  matching sample
and so there is no evidence that the incompleteness
has a spatial imprint.
\begin{figure}
\includegraphics[width=0.47\textwidth]{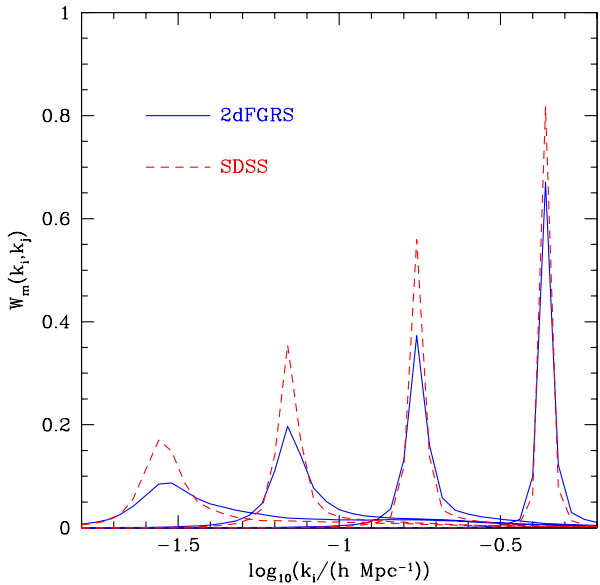}
\caption{
A sample of the normalized window functions for both SDSS (dashed)
and 2dFGRS (solid). Each curve shows the relative contribution from the underlying power
spectrum $P(k_i)$ as a function of $k_i$ to our individual band power
estimates $\hat P( k_j )$.
}
\label{fig:wins}
\end{figure}

\section{Comparison of power spectra}
\label{sec:pkcomparison}

The power spectra we estimate using the methods outlined 
in Section~\ref{sec:method} are the direct transform of the data, and are thus what CMB
researchers would term a pseudo-spectrum. As such, it yields a convolution
of the underlying galaxy power spectrum with the modulus
squared of the Fourier transform of the window function of
either the SDSS or 2dFGRS as appropriate, that is
\begin{equation}
\hat P({\bf k}) = P({\bf k}) \otimes W^2({\bf k}) .
\end{equation}
Our random catalogues allow us to accurately estimate 
$W^2({\bf k})$ and from it determine the matrix of window
functions that describe how our spherically averaged band 
power estimates $\hat P( k_j )$ are related to the unconvolved power spectrum $P( k_i )$.
\begin{equation}
\hat P( k_j ) = \sum_i P( k_i) W_{\rm m}(k_i,k_j) .
\end{equation}
Examples of these window functions for the SDSS and 2dFGRS
are shown in Fig.~\ref{fig:wins}.
For all our quantitative analysis in Section~\ref{sec:pkconstraints} we use these window functions
to convolve the model power spectra before comparing with the
data. However for the purposes of visually comparing the 
2dFGRS and SDSS power spectra we have corrected the convolved
estimates by multiplying them through by the ratio of a similar
model power spectrum to its convolved counterpart. This
`deconvolution' is accurate provided the power spectra are smooth.

\begin{figure}
\includegraphics[width=0.47\textwidth]{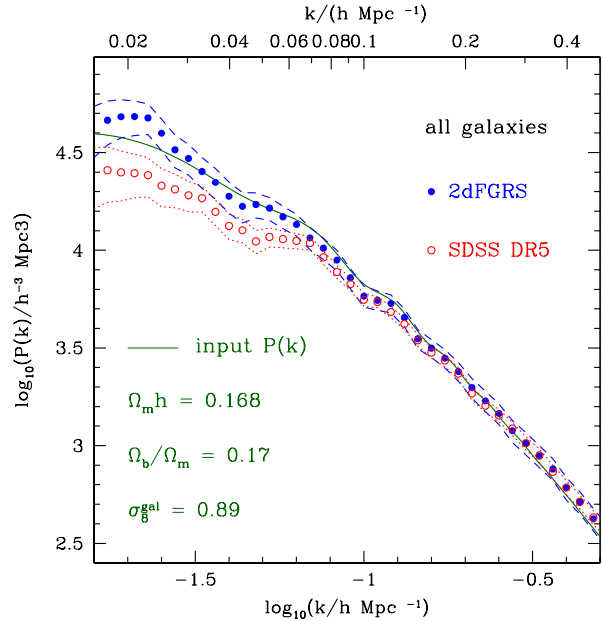}
\caption{
Comparison of the power spectra estimated from
the full 2dFGRS and SDSS DR5 samples - corrected for
the effect of the window function as described in the text. 
The solid line shows the input power spectrum of the mock catalogues
used to estimate the covariance matrix of the measurements.
}
\label{fig:pk_all}
\end{figure}

In Fig.~\ref{fig:pk_all} we compare the 
`deconvolved' power spectra estimated
from the full 2dFGRS catalogue and full SDSS-DR5 sample. 
We note that the SDSS and 2dFGRS galaxy
power spectra agree well for $k > 0.07\,h\,{\mathrm{Mpc}}^{-1}$. The good
agreement in amplitude at this wavenumber is due to the  bias
dependent weights used in the PVP estimator which have successfully
modelled the difference in the clustering strength of the red selected
SDSS galaxies and blue selected 2dFGRS galaxies. This is by design
as the bias factors were normalized empirically by the 2dFGRS red
and blue samples at this scale \citep[see figure~15 of][]{cole05}.
In contrast, on larger scales we see evidence for significantly
more large scale power in the 2dFGRS than in SDSS.
This clearly shows that the difference between the SDSS and 2dFGRS results that was
noted in the introduction and which motivated this analysis
is certainly significant and not an artifact of differing
analysis techniques.

\begin{figure}
\includegraphics[width=0.47\textwidth]{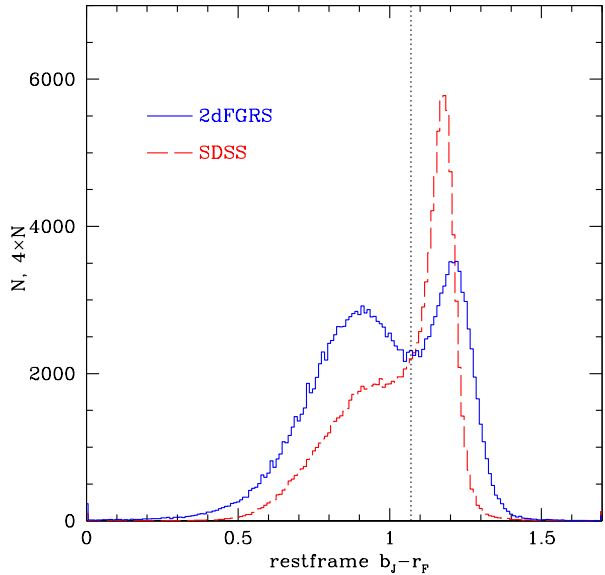}
\caption{
Histograms comparing the distribution of rest frame
$b_{\rm J} - r_{\rm F}$ colours in the 2dFGRS (solid line) and SDSS (dashed line)
catalogues. The distributions are clearly bimodal with a dividing point at 
$b_{\rm J} - r_{\rm F}=1.07$. The 2dFGRS has roughly equal numbers
of red and blue galaxies while the SDSS is dominated by red galaxies. 
Note the different units on the $y$-axis.
}
\label{fig:colhist}
\end{figure}

We now investigate if the discrepant shapes of the galaxy power
spectra are due to the difference in the clustering properties
of red and blue galaxies. Fig.~\ref{fig:colhist} shows histograms
of rest frame  $b_{\rm J} - r_{\rm F}$ colours for both the
2dFGRS and SDSS catalogues. The SDSS magnitudes have
been converted to these bands assuming the relations given
in equation~(\ref{eqn:col}). The colour distributions are clearly
bimodal with a natural dividing point at 
$b_{\rm J} - r_{\rm F}=1.07$. The 2dFGRS has roughly equal numbers
of red and blue galaxies while the SDSS, being red selected, is
naturally dominated by red galaxies. 

\begin{figure}
\includegraphics[width=0.47\textwidth]{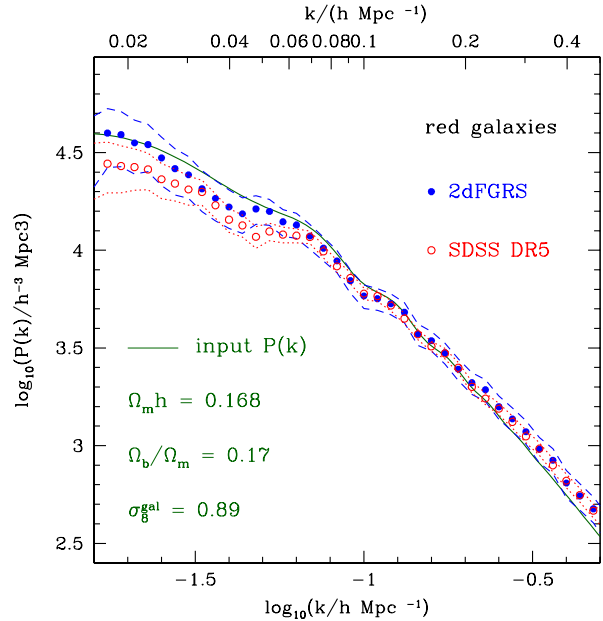}
\caption{
Comparison of the 2dFGRS and SDSS-DR5 power spectra 
from red subsamples that satisfy the rest frame colour
$b_{\rm J}-r_{\rm F}>1.07$.
The solid line shows the input power spectrum of the mock catalogues
used to estimate the covariance matrix of the measurements.
}
\label{fig:pk_red}
\end{figure}

In Fig.~\ref{fig:pk_red} we compare the galaxy power spectra estimated
from just the galaxies redder than $b_{\rm J}-r_{\rm F}=1.07$ in both
samples. Comparing the SDSS $P(k)$ from just the red galaxies with 
the previous
estimate from the full SDSS catalogue reveals them to be in very
close agreement. This is to be expected as the SDSS sample is both
dominated by red galaxies and the PVP power spectrum estimator
that we employ gives them more weight than their less clustered
blue counterparts. In contrast, the estimate for just the red
2dFGRS galaxies differs from that from all the 2dFGRS catalogue.
In fact it is a much closer match to the result from the SDSS data.
The only places where the two estimates are not in excellent
agreement is on the very largest scales $k<0.025\,h\,{\mathrm{Mpc}}^{-1}$,
where the estimates are both noisy and highly correlated,
and also around $k\approx0.05\,h\,{\mathrm{Mpc}}^{-1}$.  In fact this difference
is also due to sample variance. \citet{cole05} investigated
the effect of removing from the 2dFGRS catalogue the two
largest super clusters found by the analysis of \citet{baugh04}. Their figure~17
(panels~o and~p) shows
that this in general has a small effect, but does perturb the
power just around $k\approx0.05\,h\,{\mathrm{Mpc}}^{-1}$.

\section{Implications for cosmological parameters}
\label{sec:pkconstraints}

\subsection{The shape of the power spectrum}
\label{sec:models}

The power spectrum measured for galaxies differs in a number of ways from the power spectrum
for the mass predicted in linear perturbation theory: 
(i) Nonlinear evolution of density perturbations leads to coupling between
Fourier modes, changing the shape of the power spectrum. (ii) The
galaxy power spectrum is distorted by the gravitationally induced
peculiar motions of galaxies when a redshift is used to infer the
distance to each galaxy. (iii) The power spectrum of the galaxies
is a modified version of the power spectrum of the mass.
This phenomenon is known as galaxy bias. As we have shown in the 
previous section these effects change with scale and depend on the galaxy type.

In order to constrain cosmological parameters, these effects need
to be modelled. \citet[]{cole05} developed an empirical scheme to correct for these effects by
applying a correction for non-linearity and scale-dependent
bias to the shape of $P(k)$ of the form
\begin{equation}
P_{\rm gal}(k) = b^{2}\frac{1+Qk^{2}}{1+Ak}\,P_{\rm lin}(k),
\label{eq:nonlin}
\end{equation}
where $P_{\rm lin}(k)$ is the linear theory power spectrum and $b$ is a constant bias factor.
This formula was deduced by comparison with detailed numerical
galaxy-formation models: these show that the value of $A=1.4$ is robust, but
the exact value of $Q$ depends on galaxy type and also has some
uncertainty depending on how the modelling is done. These results were
used to determine a range of allowed values for $Q$. For 2dFGRS the value
$Q=4.6$ is preferred. 

Although this correction factor was designed and tested for the redshift space power spectrum of 2dFGRS where the scale dependent correction is small, it has also been used to model the real space power spectra for redder and more luminous galaxy samples \citep{tegmark04,tegmark06,percival07a,percival07b}. In these cases larger values of $Q$ were required to reconcile the 
measured power spectrum with linear theory \citep[e.g $Q=26$ in][]{percival07b}.
Here we analyse the validity of this correction in general terms, testing if the different shapes of the power spectra for the full and red galaxy samples of 2dFGRS and SDSS can be accounted for by using equation~(\ref{eq:nonlin}) with different values of $Q$.

Besides the change in the overall shape of $P(k)$, it has been 
shown that non-linear evolution distorts the acoustic oscillations 
\citep{eisenstein05,springel05,eisenstein06,angulo05,angulo07}. These oscillations are damped in 
a way that erases the higher harmonic peaks. \citet{eisenstein06} modelled the
damping of the oscillations by computing a `dewiggled' power spectrum by 
\begin{equation}
P_{\rm dw}(k)=P_{\rm lin}(k)G(k)+P_{\rm nw}(k)(1-G(k)),
\label{eq:bar}
\end{equation}
where $P_{\rm nw}(k)$ is a smooth 
power spectrum, with the same shape as $P_{\rm lin}(k)$ but without baryonic oscillations,
computed using the fitting formulas of \citet[]{EH99} and $G(k)\equiv \exp\left[-(k/\sqrt{2}k_{\star})^2\right]$.
This function regulates the transition from the large scales, where $P_{\rm dw}(k)$ follows linear theory to the small scales where the acoustic oscillations are completely damped. As described in \cite{tegmark06}, the value of the damping scale $k_{\star}$ is a function of $\om$ and the 
primordial amplitude of scalar fluctuations $A_{\rm s}$. 
Here we analyse the effect of this modelling on the obtained constraints on cosmological parameters.
When this correction is applied, $P_{\rm dw}(k)$ is used in equation~(\ref{eq:nonlin}) instead of $P_{\rm lin}(k)$.

We use the data from the observed power spectra for $k < 0.2\,h\,{\mathrm{Mpc}}^{-1}$ and discard 
measurements with $k < 0.02\,h\,{\mathrm{Mpc}}^{-1}$ which could be affected by uncertainties in the mean density
of galaxies within the survey. We compare the data against a restricted parameter space given by
\begin{equation}
\mathbf{
P \equiv (}
\omh,
\omb/\om,
Q,A_{\rm s}),  \label{eq:params}
\end{equation}
and we fix the values of $h=0.72$ and $n_{\mathrm{s}}=1$.
Eq.~(\ref{eq:params}) gives the most important parameters to characterize the full shape
of $P(k)$. When constraining the values of these parameters, we only use
information from the shape of the galaxy power spectrum and not from its
amplitude. This is why $A_s$ only enters in the constraints marginally when
the damping of the acoustic oscillations is included in the modelling through the assumed dependency of $k_{\star}$ on $\om$ and $A_s$ \citep{eisenstein06,tegmark06}.

\begin{figure}
\includegraphics[width=0.47\textwidth]{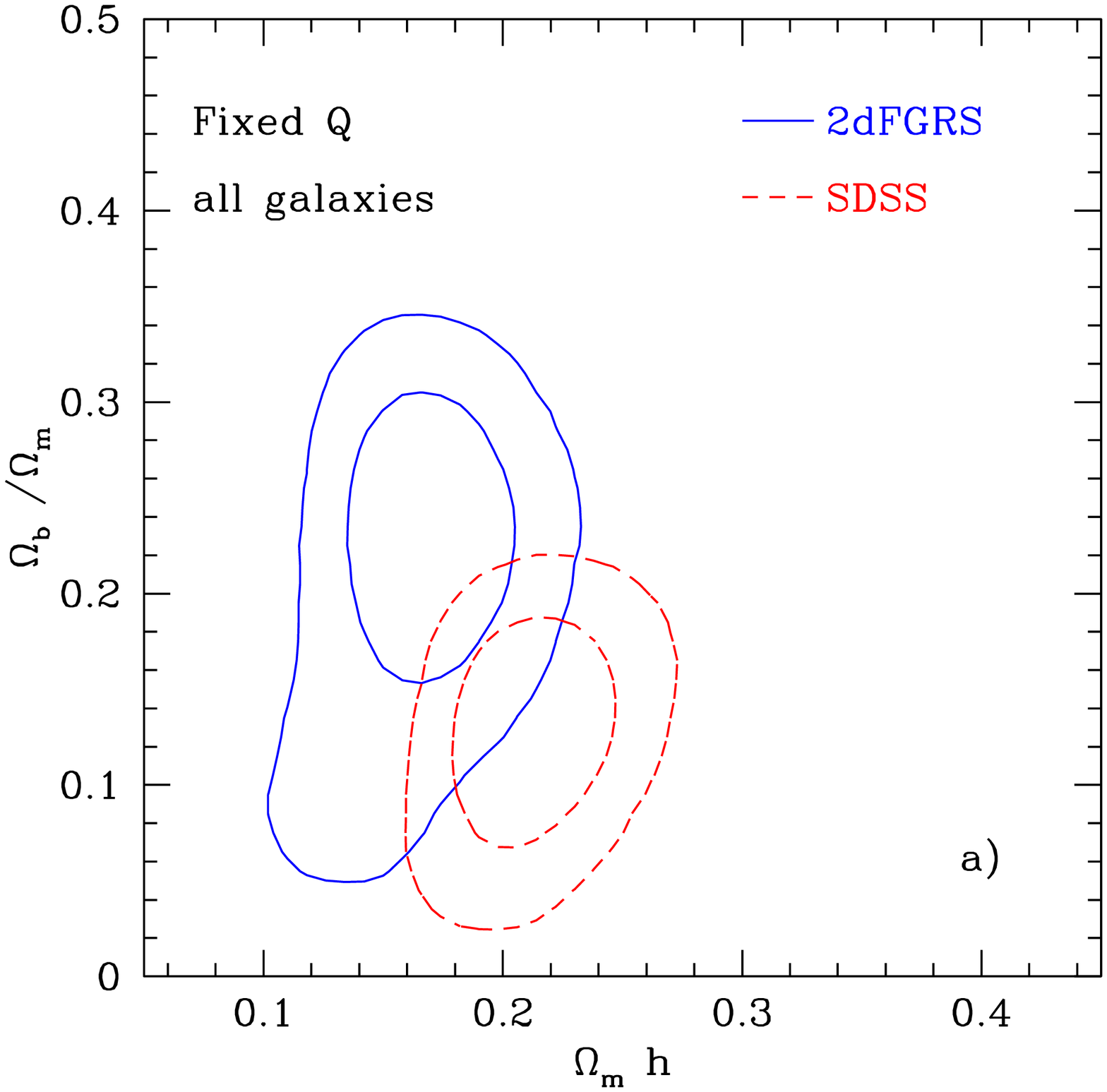}
\includegraphics[width=0.47\textwidth]{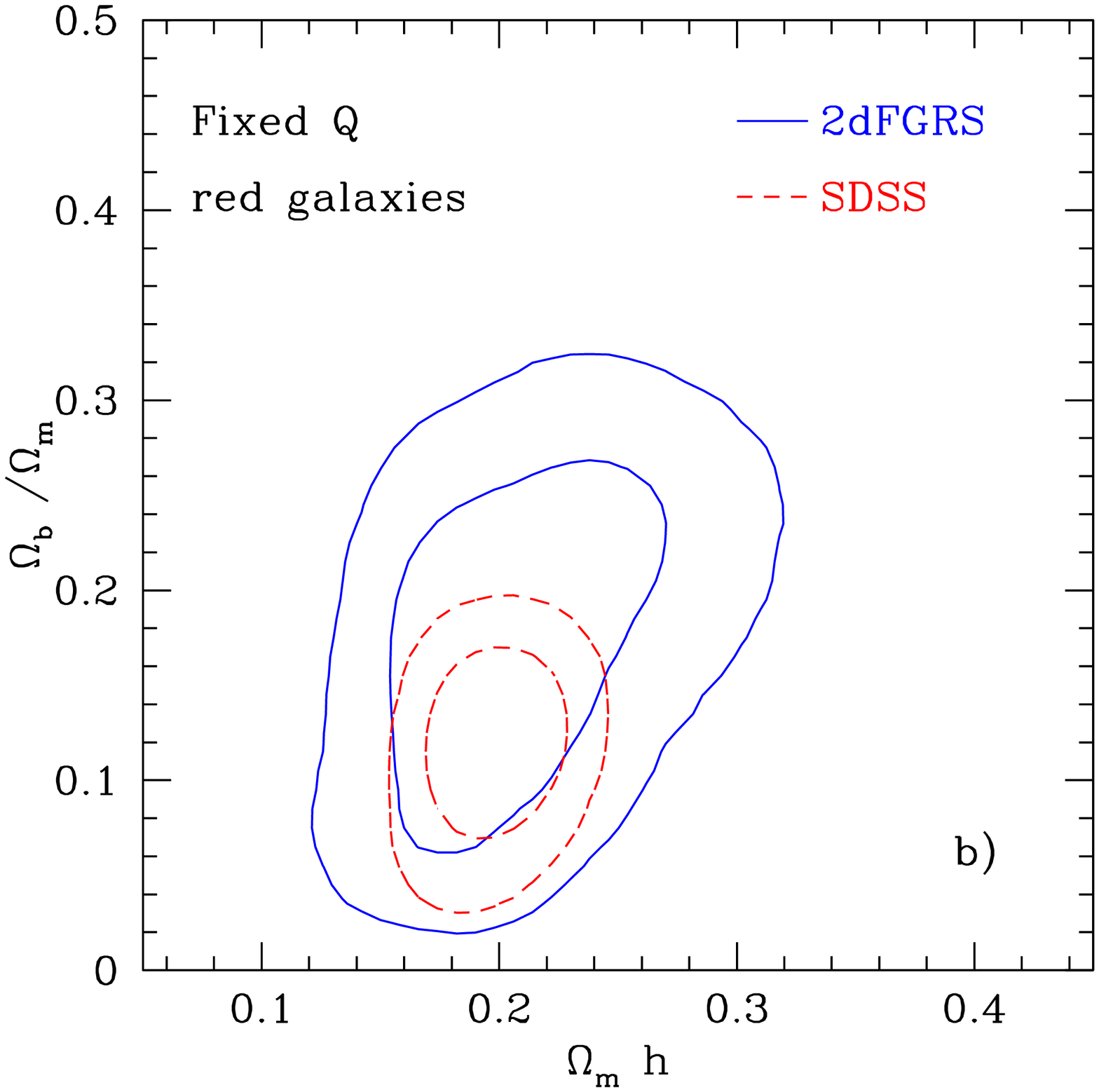}
\caption{
Contours show 68\% and 95\%
joint confidence intervals for the baryon fraction
$\Omega_{\rm b}/\Omega_{\rm m}$ and $\Omega_{\rm m} h$
for fits to the SDSS and 2dFGRS data in the range
$0.02<k<0.2\,h\,{\mathrm{Mpc}}^{-1}$. 
The parameter $Q$
modelling the distortion of power spectrum due to nonlinearity,
redshift space distortions and scale dependent bias was kept 
fixed at $Q=5$ in these fits. 
}
\label{fig:constraints}
\end{figure}

Our results were generated with a modified version of the publicly available
CosmoMC code \citep{cosmomc} as in \citet{sanchez06}. CosmoMC
uses the CAMB package to compute the linear power spectra for the matter fluctuations \citep{camb}. 
We have found that the use of the approximate \citet{EH98} fitting formula to model 
$P_{\mathrm{lin}}(k)$ causes a small shift
in the recovered values of $\omh$. 
Our analysis was carried out
in parallel on the Cosmology Machine at Durham University. For
each parameter set considered, we ran 10 separate chains using the
Message Passing Interface (MPI) convergence criterion to stop the
chains when the \citet{GR92} statistic $R < 1.01$, which
is a significantly more stringent criterion than is usually adopted \citep{verde03,seljak05}.

To test the efficiency of the modelling to account for the differences between the datasets we analyse three separate cases: first we fixed the parameters $A$ and $Q$ at fiducial values of
$A=1.4$ and $Q=5$ and ignore the correction due to the damping of the acoustic oscillations (Section~\ref{sec:fixq}). In this way the difference in the shape of the power spectra for SDSS and 2dFGRS is completely translated into the recovered constraints on $\omh$ and $\omb/\om$. Second we extend the model allowing $Q$ to vary over a wide prior (Section~\ref{sec:varq}). In this way we can test if the ansatz of equation~(\ref{eq:nonlin}) can account for the observed differences. Finally, we include the effect of the damping of the acoustic oscillations into our modelling of $P(k)$ (Section~\ref{sec:bar}). Table~\ref{tab:params} summarizes the constraints obtained with the different datasets and parameter sets analysed. 

\subsection{Fixed Q}
\label{sec:fixq}

First we analyse the constraints on $\omh$ and $\omb/\om$ that
are obtained by fitting the SDSS and 2dFGRS data with power spectra of the form described
in the last section, with the parameters $A$ and $Q$ fixed at fiducial values of
$A=1.4$ and $Q=5$ and ignoring the correction due to the damping of the acoustic oscillations.
In this way these are the only two parameters defining the shape of the model $P(k)$ and the 
difference in shape of the fitted power spectra is completely characterized by
the regions of the explored parameter space selected by each dataset. The parameter constraints 
obtained in this way are shown in Fig.~\ref{fig:constraints} for the full (upper panel) and 
red (lower panel) 2dFGRS (solid lines) and SDSS (dashed lines) galaxy power spectra.

\begin{table*}
\caption{Marginalized 68\% interval constraints on cosmological parameters 
obtained using the different datasets and parameter sets analysed. }
\begin{center}{
\small
\begin{tabular}{ccccc}
\hline
\hline
\noalign{\smallskip}
\noalign{\smallskip}
 Data set      & Parameter & Fixed $Q$ & Varying $Q$ & Varying $Q$, dw\\
\noalign{\smallskip}
\hline
\hline
\noalign{\smallskip}
        &$\Omega_{\rm m}h$                 & $0.177_{-0.025}^{+0.022}$ & $0.165_{-0.029}^{+0.025}$  & $0.191_{-0.051}^{+0.064}$ \\
\noalign{\vspace{0.1cm}}
2dFGRS all &$\Omega_{\rm b}/\Omega_{\rm m}$& $0.195_{-0.057}^{+0.054}$ & $0.212_{-0.065}^{+0.061}$  & $0.250_{-0.080}^{+0.081}$ \\
\noalign{\vspace{0.1cm}}
        &$Q$                               & -                         & $7.7_{-3.4}^{+3.3}$&  $8.1_{-3.8}^{+3.7}$   \\
\noalign{\smallskip}
\hline
\noalign{\smallskip}
        &$\Omega_{\rm m}h$                & $0.217_{-0.040}^{+0.040}$  &  $0.197_{-0.043}^{+0.043}$  &  $0.023_{-0.069}^{+0.094}$ \\
\noalign{\vspace{0.1cm}}
2dFGRS red &$\Omega_{\rm b}/\Omega_{\rm m}$  & $0.183_{-0.068}^{+0.064}$ & $0.197_{-0.074}^{+0.071}$ &  $0.238_{-0.089}^{+0.086}$ \\
\noalign{\vspace{0.1cm}}
        &$Q$                              & -                          & $8.0_{-4.5}^{+4.4}$&  $8.1_{-4.7}^{+4.8}$   \\
\noalign{\smallskip}
\hline
\noalign{\smallskip}
         &$\Omega_{\rm m}h$              & $0.199_{-0.014}^{+0.014}$ & $0.216_{-0.020}^{+0.019}$ & $0.234_{-0.034}^{+0.030}$ \\
\noalign{\vspace{0.1cm}}
SDSS all &$\Omega_{\rm b}/\Omega_{\rm m}$& $0.128_{-0.031}^{+0.029}$ & $0.112_{-0.033}^{+0.019}$ & $0.150_{-0.053}^{+0.050} $ \\
\noalign{\vspace{0.1cm}}
         &$Q$                            & -                         & $2.8_{-1.5}^{+1.5}$       & $3.1_{-1.6}^{+1.6}$   \\
\noalign{\smallskip}
\hline
\noalign{\smallskip}
        &$\Omega_{\rm m}h$                & $0.201_{-0.015}^{+0.013}$ & $0.221_{-0.025}^{+0.017}$ & $0.240_{-0.039}^{+0.038}$ \\
\noalign{\vspace{0.1cm}}
SDSS red &$\Omega_{\rm b}/\Omega_{\rm m}$ & $0.120_{-0.030}^{+0.027}$ & $0.115_{-0.036}^{+0.031}$ & $0.157_{-0.058}^{+0.057}$ \\
\noalign{\vspace{0.1cm}}
        &$Q$                              & -                         & $3.2_{-1.7}^{+1.6}$       & $3.5_{-1.7}^{+1.7}$   \\
\noalign{\smallskip}
\hline
\noalign{\smallskip}
        &$\Omega_{\rm m}h$                & -                 & $0.172_{-0.016}^{+0.016}$ &  $0.173_{-0.017}^{+0.017}$ \\
\noalign{\vspace{0.1cm}}
LRG     &$\Omega_{\rm b}/\Omega_{\rm m}$  & -                 & $0.173_{-0.036}^{+0.038}$ &  $0.187_{-0.038}^{+0.038}$ \\
\noalign{\vspace{0.1cm}}
        &$Q$                              & -                 & $27.1_{-4.7}^{+4.7}$      &  $28.5_{-5.0}^{+4.9}$   \\
\noalign{\smallskip}
\hline
\hline
\end{tabular}
}\end{center}
\label{tab:params}
\end{table*}

Fig.~\ref{fig:constraints}(a) shows that the 2dFGRS constraints are in 
complete agreement with those obtained by CMB data \citep{sanchez06,spergel07}.
For 2dFGRS data only we obtain $\Omega_m=0.246 \pm 0.030$ while the WMAP3 analysis of \citet{spergel07} gives $\Omega_m=0.234 \pm 0.035$.
The SDSS parameter estimates are completely in accord with
those from \citet{tegmark04}, but with much tighter bounds due to the larger SDSS dataset used here.
This shows the consistency of the different analysis techniques applied to this dataset.

We note that the 2dFGRS and SDSS best fit values lie
outside each others 95\% confidence contours.
 This clearly shows that, if not accounted for, the difference between the SDSS and
2dFGRS power spectra that was noted in Section~\ref{sec:pkcomparison} causes a bias in the
obtained constraints on cosmological parameters which is larger than the uncertainty with
which they are determined.

Fig.~\ref{fig:constraints}(b) confirms the agreement, within the expected 
statistical uncertainty, of the power spectra for red galaxies in 2dFGRS and SDSS. Being dominated by red galaxies,
the SDSS constraints are almost identical to the ones obtained with the full sample. Instead, the results for 2dFGRS
shift towards the same region of the parameter space preferred by SDSS to contain it completely. The best
fitting parameters for 2dFGRS lie within the SDSS 65\% confidence contour and vice versa.
The contours obtained for 2dFGRS broaden as a result of the larger errors in the power spectrum obtained with a smaller sample of galaxies. This shows that the differences in the constraints obtained with SDSS and 2dFGRS are due to the different selection criteria used to construct the samples, with the $r$-band selected SDSS catalogue being dominated by
more strongly clustered red galaxies which are more strongly affected by scale dependent bias.

\begin{figure*}
\begin{center}
$\begin{array}{cc}
\multicolumn{1}{l}{} &
\multicolumn{1}{l}{} \\ [-0.53cm]
\epsfxsize=0.47\textwidth
\epsffile{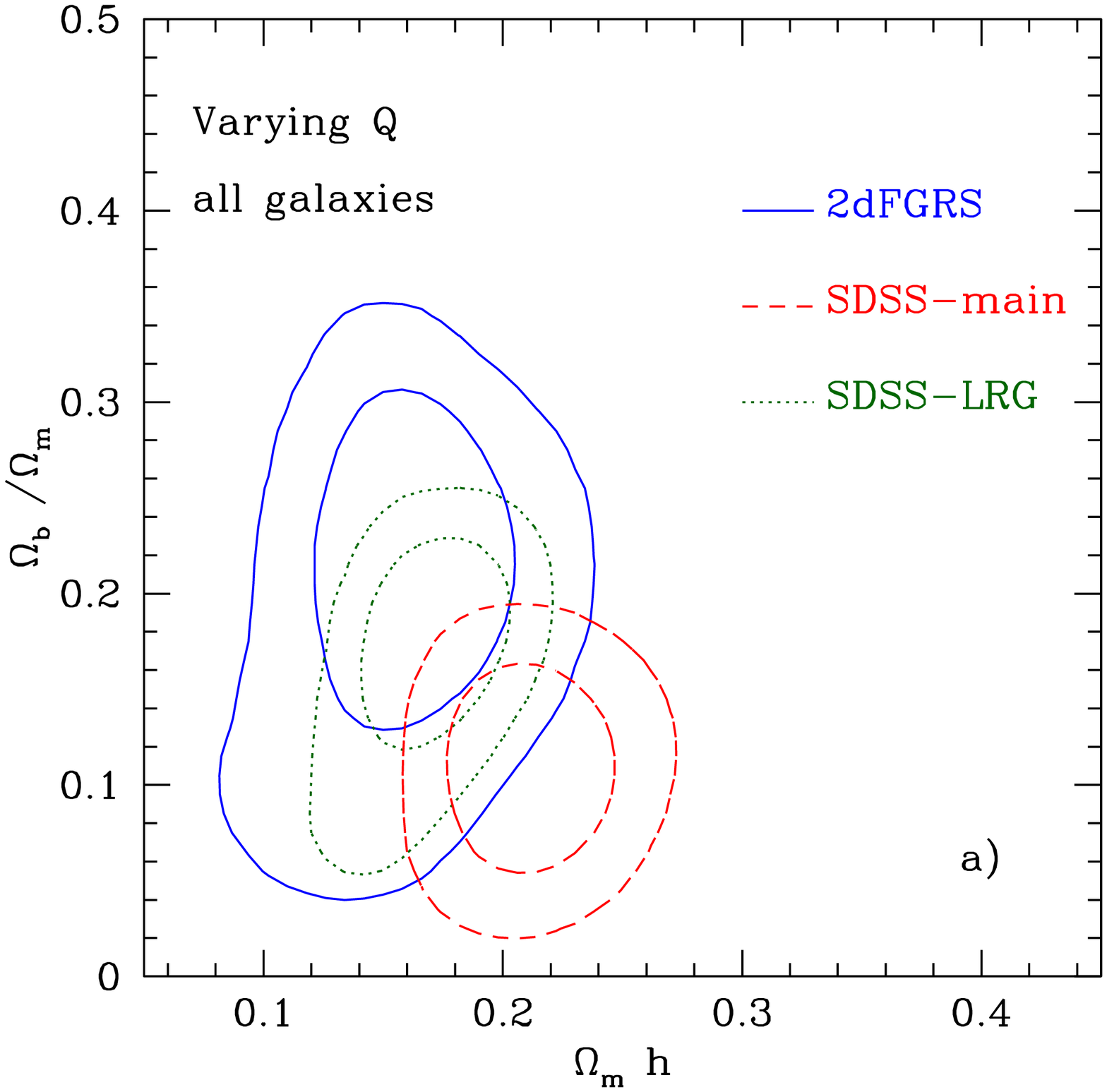} &
\epsfxsize=0.47\textwidth
\epsffile{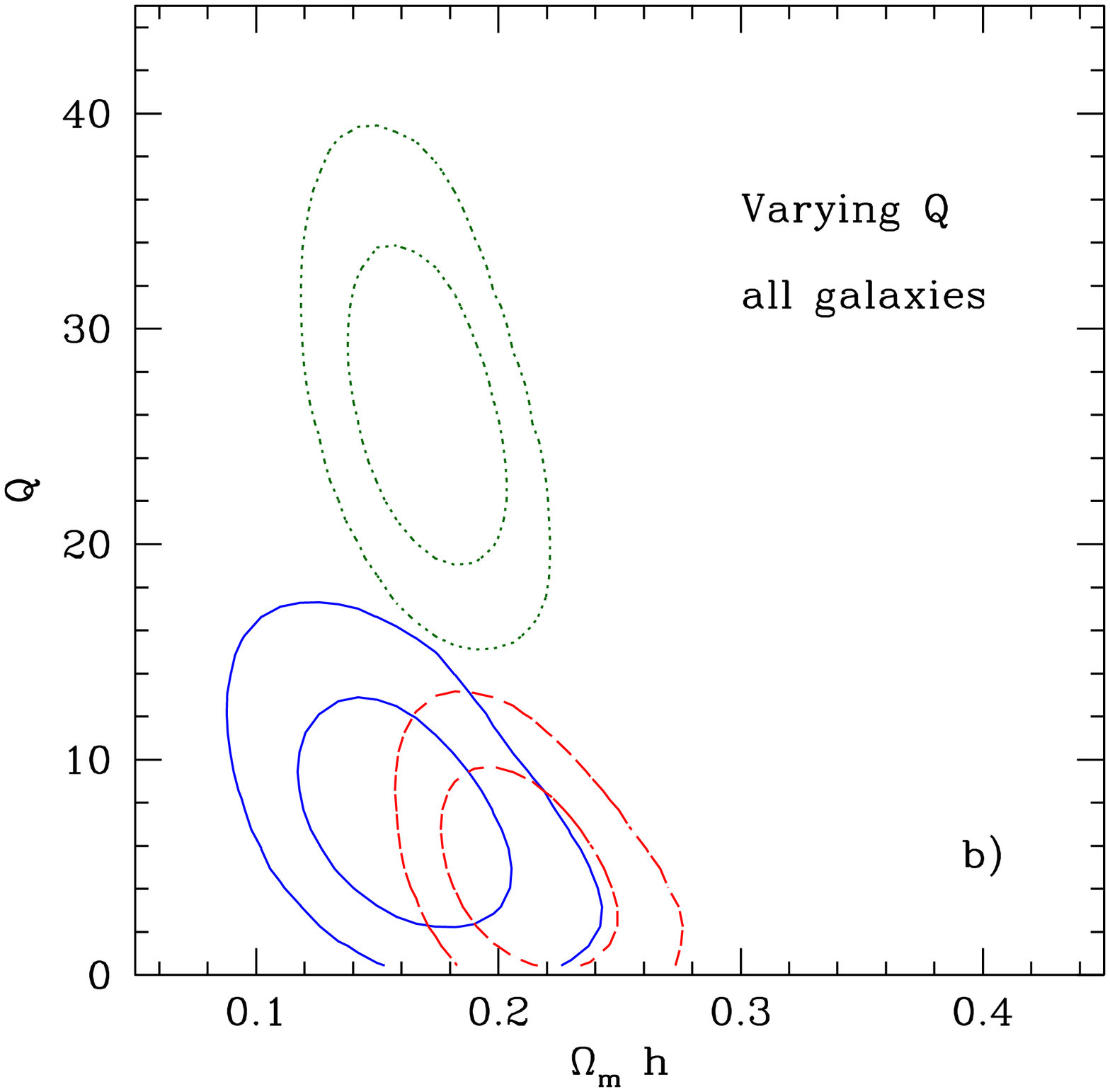} \\ [0.4cm]
\end{array}$

$\begin{array}{cc}
\multicolumn{1}{l}{} &
\multicolumn{1}{l}{} \\ [-0.53cm]
\epsfxsize=0.47\textwidth
\epsffile{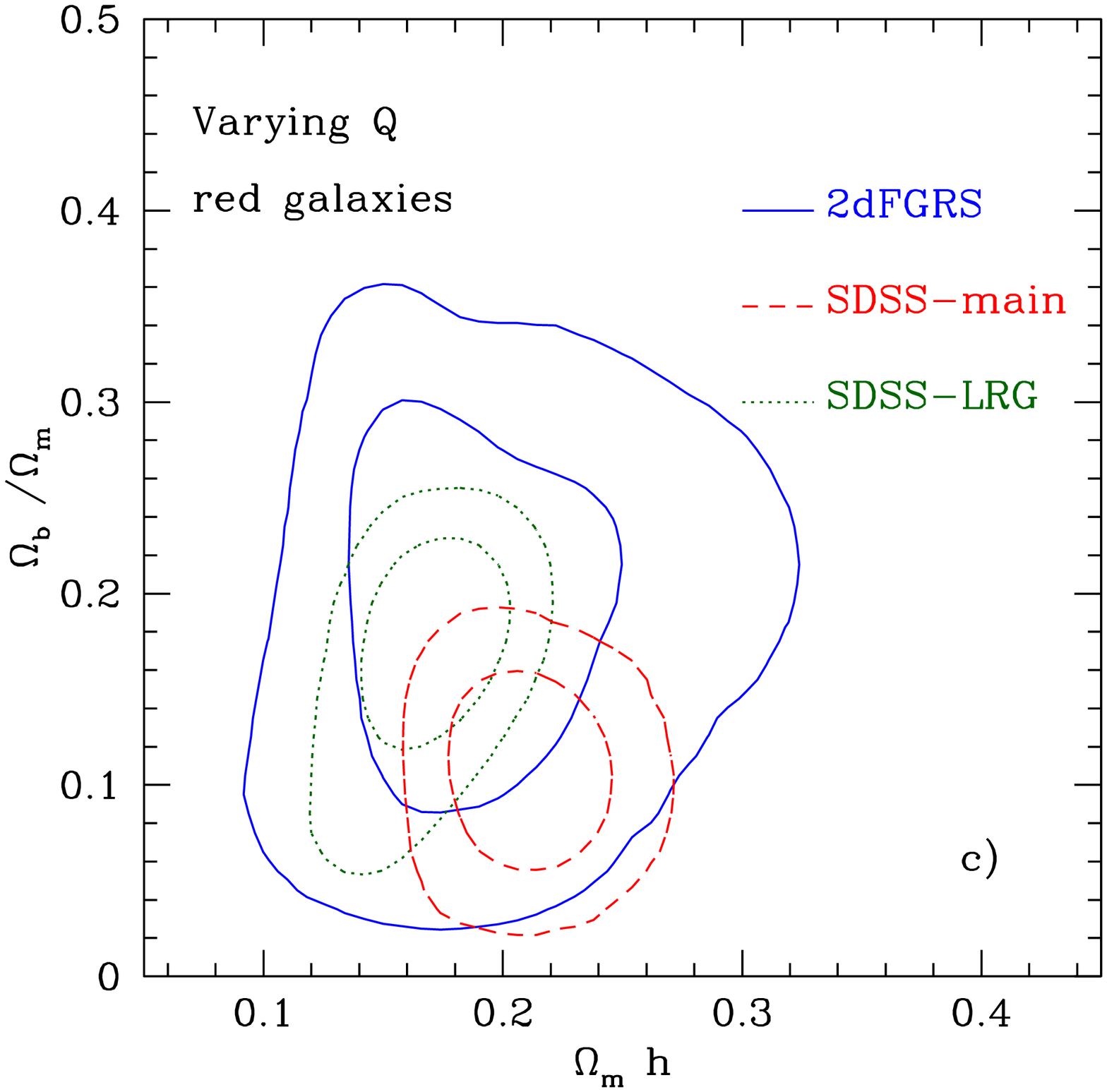} &
\epsfxsize=0.47\textwidth
\epsffile{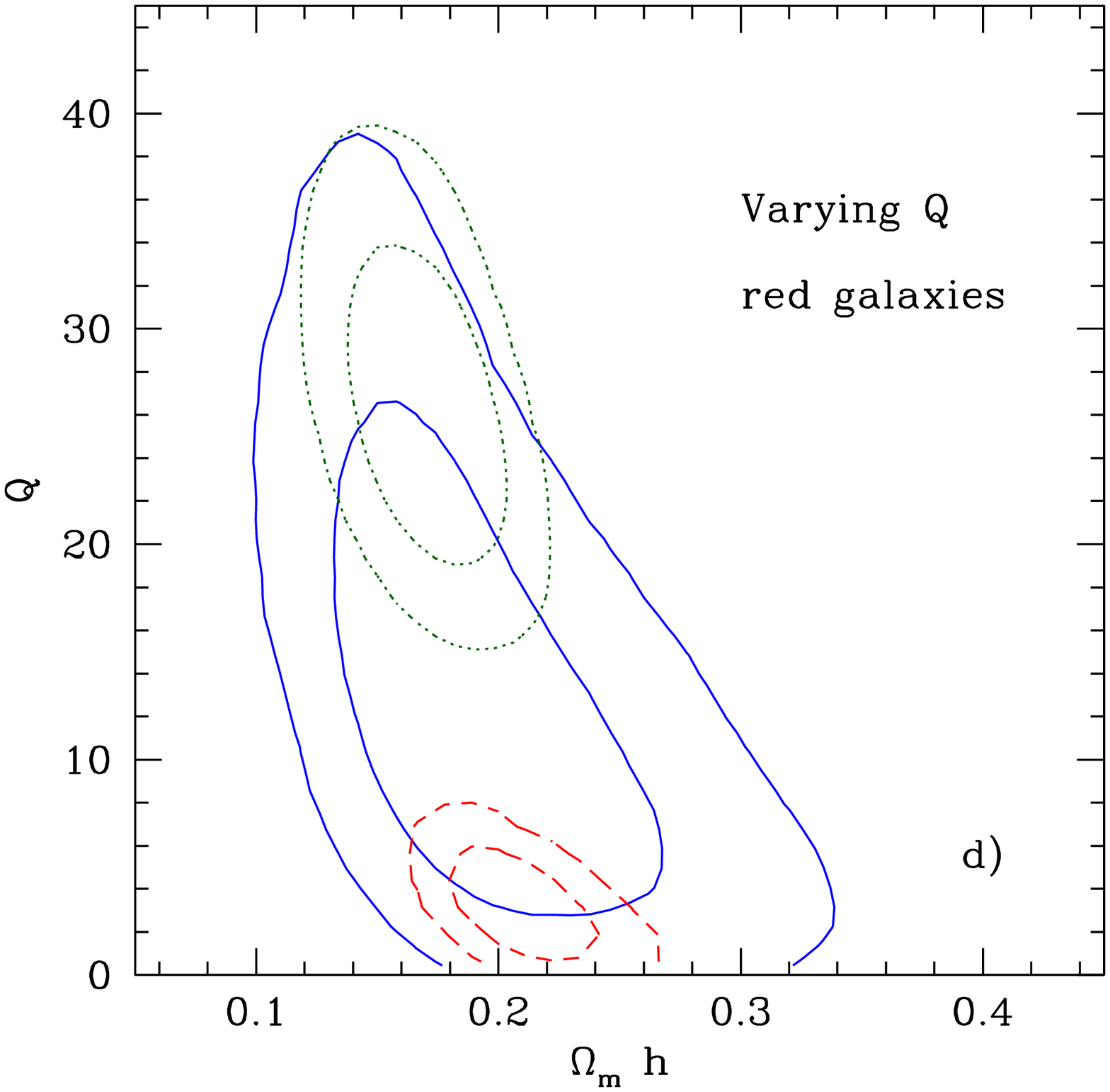} \\ [0.4cm]
\end{array}$
\end{center}
\caption{ 
Contours show 68\% and 95\%
joint confidence intervals for the baryon fraction
$\Omega_{\rm b}/\Omega_{\rm m}$ and $\Omega_{\rm m} h$
for fits to the 2dFGRS (solid line), SDSS (dashed line) and SDSS-LRG (dotted line) in the range
$0.02<k<0.2\,h\,{\mathrm{Mpc}}^{-1}$. 
In these fits, the parameter $Q$
modelling the distortion of power spectrum due to nonlinearity,
redshift space distortions and scale dependent bias was allowed to vary over a wide prior $(0-45)$.
}
\label{fig:varq}
\end{figure*}

\subsection{Varying Q}
\label{sec:varq}

In this section we extend the parameter space by allowing $Q$ to vary over a wide prior.
This allows us to test if the shape distortion modelled by equation~(\ref{eq:nonlin}) can account
for the differences between 2dFGRS and SDSS found in Section~\ref{sec:pkcomparison}. 
The results obtained in this way are shown in Fig.~\ref{fig:varq}
for the full (upper panels) and red (lower panels) 2dFGRS (solid lines) and SDSS (dashed lines)
galaxy power spectra. For this case, we have also included  in our analysis the power spectrum of the SDSS 
Luminous Red Galaxies (LRG, dotted lines in Fig.~\ref{fig:varq}) as measured by
\citet{tegmark06}.

From Figs.~\ref{fig:varq}(a) and (b) we note that the
correction factor of equation~(\ref{eq:nonlin}) is not able to reconcile
the constraints obtained from the 2dFGRS and SDSS power spectra. The
constraints on the $\omh - \omb/\om$ plane exhibit a similar behaviour to the 
case where the value of $Q$ was held fixed. This shows that the non-linearities
and scale dependent bias of the $r$-selected SDSS galaxies distort the shape of the power
spectrum in a way that can not be described accurately by equation~(\ref{eq:nonlin}). 
Contrary to what was expected, the constraints on the plane $\omh - Q$ show that SDSS
and 2dFGRS select different values of $\omh$ but a similar allowed region for $Q$. 
Then, even incorporating this model of the distortion into the methodology for constraining
cosmological parameters, the use of a red sample of galaxies introduces strong systematics
effects on the resulting parameter constraints.

Intriguingly, the constraints on $\omh$ and $\omb/\om$ obtained using the LRG power 
spectrum lie well within the 2dFGRS contours, in perfect agreement with the results obtained
using CMB data. This means that the difference in the shape of the SDSS-LRG and 2dFGRS power
spectra is completely consistent with equation~(\ref{eq:nonlin}). Both datasets can be
accurately described with the same values of the cosmological parameters, but with a much
higher value of $Q=27.1\pm 4.7$ for the SDSS-LRG sample, in agreement with the results
of \citet{percival07b}.

\begin{figure*}
\includegraphics[width=0.8\textwidth]{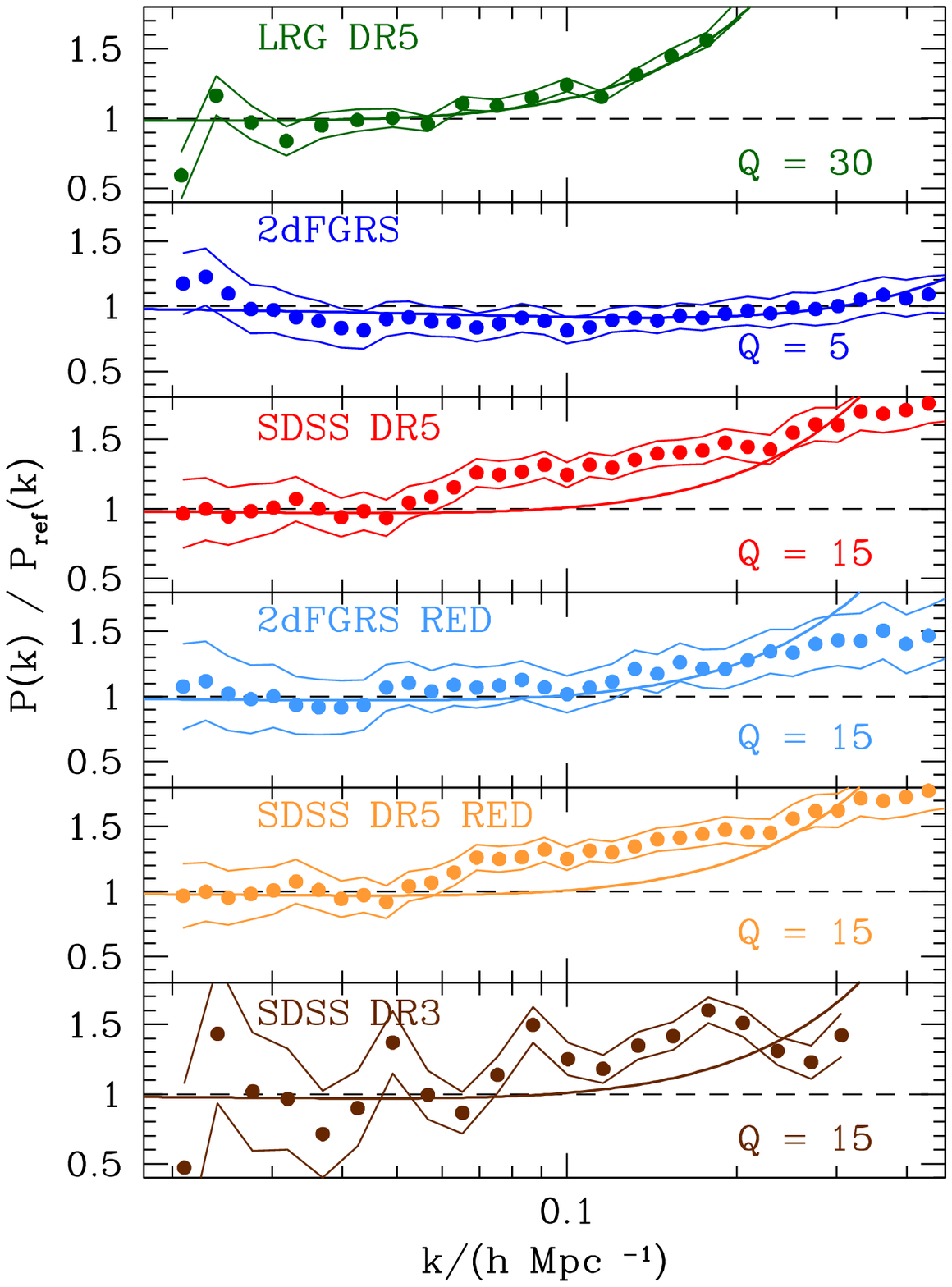}
\caption{
The power spectra analysed in this work 
divided by a reference power spectrum computed using the best fit cosmological 
parameters from \citet{spergel07}. The solid lines show the shape distortion modelled by equation~(\ref{eq:nonlin}) for the specified value of $Q$.
}
\label{fig:ratio}
\end{figure*}

As shown in Fig.~\ref{fig:varq}(c) and (d), the situation is similar
for the constraints obtained using the power spectra of the red 2dFGRS and SDSS
galaxy samples. As in the case of a fixed value of $Q$, the constraints obtained
with the 2dFGRS power spectrum are shifted from its region of agreement with the
CMB towards the values preferred by the SDSS. With the addition of $Q$, the extra
degree of freedom gives rise to a degeneracy that involves $\omh$ and $Q$. These 
two parameters act in opposite ways which allows them to compensate each others
effect on the theoretical power spectrum.
This degeneracy can not be broken by the power spectrum of the red 2dFGRS galaxies
and results in a considerable increase of the allowed region for the different parameters.

The breakdown of the simple model of equation~(\ref{eq:nonlin}) in describing the 
differences in the clustering of red and blue galaxies can be more clearly seen in
Fig.~\ref{fig:ratio}, which shows the different power spectra analysed in this work 
divided by a reference power spectrum $P_{\mathrm{ref}}(k)$ computed using the best fit cosmological 
parameters from \citet{spergel07}. Assuming this is the true cosmology, if there were 
no distortions in the shape of $P(k)$, the different measurements should lie over the
dashed line of $P(k)/P_{\mathrm{ref}}(k)=1$. The difference in the shapes of these spectra is a 
clear demonstration of the fact that scale dependent bias and nonlinearities are a
function of galaxy type. The solid lines in Fig.~\ref{fig:ratio} show the shape distortion model of equation~(\ref{eq:nonlin}) for the specified value of $Q$.
With a value of $Q\approx5$ this correction gives a good description of the distortions
in the power spectrum of the 2dFGRS, with a small decrease in the amplitude of $P(k)$ at intermediate
scales and an increase at small scales.
The same correction factor can reproduce the effect of the larger non-linearities of the
LRG power spectrum, but with a much higher value of $Q\approx30$.

Despite its success at describing the 2dFGRS and LRG power spectra, equation~(\ref{eq:nonlin}) fails to reproduce the shape of the SDSS power spectrum, which exhibits a different behavior as a function of $k$ and can not be fitted by this simple model. The fact that the power spectrum of the red galaxies from 2dFGRS is distorted in a similar fashion indicates that, as was found in Section~\ref{sec:pkcomparison}, red selected galaxy samples exhibit stronger scale dependent galaxy bias. For comparison, Fig.~\ref{fig:ratio} also shows that the power spectrum of SDSS DR3 measured by \cite{tegmark04} suffers from the same effect. 

This clearly shows that, in order to obtain unbiased constraints on cosmological parameters
it is necessary to better understand and constrain the distortion in the shape
of the power spectrum caused by nonlinearity and scale dependent bias.

\subsection{Damping of acoustic oscillations}
\label{sec:bar}

In this section we analyse the effect of including the damping of the acoustic
oscillations into our modelling of $P(k)$. This case mimics the analysis of
\citet{tegmark06} of the LRG power spectrum. 
Although we only use information from the shape of the galaxy power spectrum,
when this effect is included in our modelling of $P(k)$ the amplitude $A_s$
enters in the constraints marginally, since the value of the damping scale 
$k_{\star}$ depends on this parameter \citep{eisenstein06,tegmark06}.

As can be seen in the last column
of Table~\ref{tab:params} this addition does not change the differences between
2dFGRS and SDSS. The main effect of this is the increase of the allowed regions
of the different cosmological parameters. As the oscillations are damped, models 
with higher values of $\omb/\om$, which according to linear theory exhibit larger
amplitude oscillations, are now compatible with the data. This causes the 
signal of the oscillations to effectively loose the power to constrain 
efficiently the values of the parameters, which in this case are determined almost
completely by the overall shape of $P(k)$.

The details of the way in which the oscillations are damped and the correct way to
model it are the main focus of several recent studies \citep{eisenstein06,angulo07,crocce07}.
Although substantial progress has been made in this subject, it is not completely understood. 
This is then another example of how the details of the modelling used to fit theoretical 
models may introduce systematic effects which may cause significant differences in the 
obtained constraints.

\section{Conclusions}   
\label{sec:conc}

In this paper we studied in detail the differences between the
constraints on cosmological parameters coming from the 2dFGRS and SDSS
power spectra. We have analysed both the 2dFGRS 
and SDSS DR5 catalogues using essentially identical techniques.
The main conclusion of this investigation is that a significant difference
exists between the shape of the galaxy power spectra measured in the
2dFGRS and SDSS main galaxy sample and that this difference is due to scale
dependent bias.
Our results extend those of \citet{percival07a} who found that the inclusion
of the LRGs in the analysis increases the discrepancy between 2dFGRS and SDSS.
Their findings also showed that if linear CDM fits are restricted to large scales ($0.01<k<0.06\,h~\mathrm{Mpc}^{-1}$), SDSS data also favours
a lower matter density $\Omega_mh=0.16 \pm 0.03 $.
\citet{percival07a} suggested that a scale-dependent bias which depends on the $r$-band
luminosity of the galaxies could explain these differences, but found no significant
evidence of it.

If a homogeneous sample of red galaxies is selected
from each survey then the resulting power spectra agree to within
the expected statistical errors.
In contrast, when the full 2dFGRS and SDSS catalogues are analyzed the 
resulting 2dFGRS power spectrum differs in shape to that of SDSS.
If normalized on scales around $k\approx0.1\,h\,{\mathrm{Mpc}}^{-1}$,
as is done automatically by the scale independent bias factors
assumed in the PVP estimator, then the 2dFGRS $P(k)$ exhibits more large
scale power than SDSS. However, if one instead normalizes on large
scales one finds the equivalent result that SDSS exhibits more
small scale power than 2dFGRS.
This behaviour is exactly what one expects if the more strongly
clustered red galaxies live in denser environments where the effects
of nonlinearity are greater.

We have obtained constraints on the parameter combinations $\omh$ and $\omb/\om$ which are the
most important parameters to characterize the full shape of $P(k)$.
The constraints obtained by fitting the SDSS and 2dFGRS data show that the difference
between the SDSS and 2dFGRS results which motivated this analysis
is significant and is not due to the differing analysis techniques.
If not accounted for, the difference between these datasets introduces a systematic effect that bias the
obtained constraints on cosmological parameters, whose effect is larger than the uncertainty with
which they are determined.

\citet{cole05} introduced a first attempt at modelling
the distortion in the shape of the power spectrum caused by nonlinearity and scale
dependent bias using the $Q$ and $A$ parameters of equation~(\ref{eq:nonlin}). 
For the mix of red and blue galaxies
present in the 2dFGRS survey, the necessary value of $Q$ was reasonably
small and hence the scale dependent correction quite
modest. The same correction factor has been applied to samples of redder
or more luminous and hence more clustered galaxies where one expects greater
nonlinearity.

We found that the correction factor of equation~(\ref{eq:nonlin}) is not able to reconcile
the constraints obtained from the 2dFGRS and SDSS power spectra. This shows that this simple correction can not describe the effect of non-linearities and scale dependent bias for a general sample of galaxies.
Intriguingly, the same model can describe correctly the difference in the shape of the power spectra from 2dFGRS and the much strongly clustered SDSS-LRG sample. Both datasets can be described with the same values of the cosmological parameters, but with a much higher value of $Q=27.1\pm 4.7$ for the SDSS-LRG sample.

This comparison has revealed that to get unbiased estimates
of the cosmological parameters it is necessary to better
understand and constrain the different processes that shape the galaxy power spectrum. 
Thus to get robust constraints from the main SDSS survey and from future 
galaxy surveys like Pan-STARRS, or the Dark Energy Survey, which will use selection
criteria similar to the one of SDSS, will require
more detailed modelling of nonlinearity and scale dependent bias.

\section*{Acknowledgements}

We thank Steve Wilkins for his assistance in preparing preliminary
versions of many of our plots.
We thank Carlton Baugh for his careful reading of the manuscript and useful discusions.
AGS acknowledges the hospitality of the Department of Physics 
at the University of Durham where part of this work was carried out.
AGS acknowledges a fellowship from CONICET;
This work was supported by PPARC, the European Commission's ALFA-II programme 
through its funding of the Latin-american European Network for Astrophysics 
and Cosmology (LENAC), and the Royal Society, through the award of an International
Incoming Short Visit grant.

\end{document}